\documentclass[superscriptaddress,prb,twocolumn]{revtex4-2}
\usepackage{graphicx}
\usepackage{amsmath,amssymb,physics,dsfont}
\usepackage{mathrsfs}
\usepackage{hyperref}
\usepackage{microtype}
\usepackage{braket}
\usepackage{defs-private}
\usepackage{booktabs}
\usepackage[dvipsnames]{xcolor}

\begin{document}

\title{Adaptive ghost Gutzwiller dynamics of Hubbard model quenches\texorpdfstring{\\}{ } across coupling strengths over extended times}

\author{Yong-Xin Yao}
\email{ykent@iastate.edu}
\affiliation{Ames National Laboratory, U.S. Department of Energy, Ames, Iowa 50011, USA}
\affiliation{Department of Physics and Astronomy, Iowa State University, Ames, Iowa 50011, USA}

\begin{abstract}
Following the real-time dynamics of correlated electrons to extended times, at all interaction strengths, remains an open challenge. We address it with an adaptive time-dependent ghost Gutzwiller approximation (td-gGA). The method extends the Gutzwiller approximation by auxiliary ghost orbitals. It adds more of them whenever the correlations that build up during the dynamics exceed a threshold. This is computationally far more efficient than keeping their number fixed. As a demonstration we take the interaction quench of the half-filled Hubbard model on the Bethe lattice. For this paradigmatic problem, numerically exact nonequilibrium dynamical mean-field theory (DMFT) serves as the reference. A previous td-gGA study with a fixed set of 6 ghost orbitals per spin largely reproduced the available DMFT data. It lacked, however, an error estimate that would support predictions beyond them. With up to 16 ghost orbitals per spin, adaptive td-gGA agrees with DMFT wherever DMFT data are available, from the weak-coupling prethermal regime to the strong-coupling collapse-and-revival regime. At strong coupling it remains converged up to three and a half times as long as the DMFT data. Near the crossover between the two regimes, the system relaxes to thermal values that agree with DMFT. Resolving the dynamics in the number of ghost orbitals shows how the sharp dynamical transition of the Gutzwiller approximation turns into this smooth crossover. In the extended time window at strong coupling, we observe that the oscillation frequency is renormalized as predicted by our second-order strong-coupling expansion. With its built-in error estimate, adaptive td-gGA offers a systematically improvable route to the extended-time dynamics of driven and multi-orbital correlated systems, especially when combined with tensor-network or quantum solvers for the embedding problem.
\end{abstract}

\maketitle

\section{Introduction}

The real-time dynamics of interacting electrons underlies a broad range of nonequilibrium phenomena. They range from the ultrafast spectroscopy and light control of correlated materials~\cite{ultrafast_approach, de2021colloquium} to the dynamics of fermions in optical lattices~\cite{Bloch2008RMP, Esslinger2010, Strohmaier2010doublon}. Describing this dynamics is a central task of many-body theory~\cite{Polkovnikov2011RMP, RMPNonequilibriumDMFT}. The task is demanding, because the state explores an exponentially large Hilbert space. Numerically exact methods are limited to small systems or short times, and approximate methods are typically controlled only in limiting regimes. Methods that reach extended times at any interaction strength, and that quantify their own accuracy, are therefore needed.

A paradigmatic problem, and a stringent benchmark for such methods, is how an isolated, interacting quantum many-body system approaches thermal equilibrium after a sudden perturbation~\cite{Polkovnikov2011RMP, Rigol2008thermalization, Rigol2007GGE}. For correlated electrons the standard protocol is the interaction quench of the Hubbard model~\cite{Hubbard1963}, prepared in the noninteracting Fermi sea and evolved with a finite interaction $U$~\cite{Moeckel08InteractionQH, Eckstein09ThermalizationIQ}. In infinite dimensions, nonequilibrium dynamical mean-field theory (DMFT) maps this problem exactly onto a self-consistent impurity problem~\cite{dmft_georges96, RMPNonequilibriumDMFT}. Its numerical solution revealed two dynamical regimes~\cite{Eckstein09ThermalizationIQ}. For weak quenches, the system is trapped in a prethermal state whose momentum distribution retains a reduced Fermi-surface discontinuity~\cite{Moeckel08InteractionQH, Kollar2011GGE}. For strong quenches, the double occupancy and the Fermi-surface discontinuity show collapse-and-revival oscillations with a period set by $U$. The two regimes are separated by a narrow crossover around $U\simeq 1.6$ in units of the half-bandwidth, where the system thermalizes fast~\cite{Eckstein09ThermalizationIQ}. The numerical solution, however, is limited to short and intermediate times once the interaction is strong. The impurity problem was solved with real-time weak-coupling continuous-time quantum Monte Carlo. Its dynamical sign problem grows exponentially with the perturbation order and thus with the interaction and the propagation time~\cite{Eckstein2010quenchspectral}. As a result, the data of Ref.~\cite{Eckstein09ThermalizationIQ} reach $t=10$ in units of the inverse half-bandwidth for $U\leq 1.5$, but only $t=2.6$ at $U=4$. Other impurity solvers for nonequilibrium DMFT have since been developed. They are based on strong-coupling expansions around the atomic limit~\cite{Eckstein2010NCA}, on inchworm quantum Monte Carlo~\cite{Dong2017inchwormDMFT}, and on tensor cross interpolation of the weak-coupling expansion~\cite{Matsuura2026TCI}. Others represent the impurity problem by a Hamiltonian with a finite bath, treated by exact diagonalization~\cite{Gramsch2013Hamiltonian}, multiconfiguration time-dependent Hartree~\cite{Balzer2015MCTDH} or matrix product states~\cite{Wolf2014MPSnoneqDMFT}. These solvers have broadened the reach of nonequilibrium DMFT to other regimes and protocols. For this particular quench from the Fermi sea, however, the numerically exact solution at strong coupling has, to our knowledge, not yet been extended to longer times. Weak-coupling perturbation theory reaches extended times and conserves the energy up to $U\simeq 1$~\cite{Tsuji2013weakcoupling}. The tensor-cross-interpolation solver reproduces the quantum Monte Carlo results over a comparable time range~\cite{Matsuura2026TCI}. Simulations to extended times over the whole range of interactions thus remain challenging.

The time-dependent Gutzwiller approximation (td-GA)~\cite{ga, Seibold2001tdGA, Oelsen2011multiband, Fabrizio2013tdGA} captures the short-time dynamics. It has a sharp dynamical transition~\cite{schiro10tdga, SchiroFabrizio2011, Sandri2012ramps} at $U_c^\mathrm{dyn}=U_c^\mathrm{BR}/2$, half the Brinkman-Rice interaction~\cite{BrinkmanRice}, close to the DMFT crossover. However, td-GA describes the dynamics by a single collective degree of freedom: observables oscillate forever and never relax~\cite{SchiroFabrizio2011}. The ghost Gutzwiller approximation (gGA) enlarges the variational space by $B-1$ auxiliary ``ghost'' orbitals per physical orbital~\cite{gga_lanata2017, Frank2021QuantumED, Lanata2022operatorial, Lee2023AccuracyGRISB, MejutoZaera2023screening, Lee2023multiorbGRISB, mejuto2024quantum}. It is equivalent to the ghost extension of the rotationally invariant slave-boson theory~\cite{sb_kotliar86, Lechermann_risb, Lanata2022operatorial} and converges rapidly to DMFT in equilibrium as $B$ grows~\cite{Lee2023AccuracyGRISB}. Its time-dependent extension (td-gGA)~\cite{Guerci2023TDgGA} damps the td-GA oscillations already for $B=3$. However, gGA maps each site onto an embedding problem: an impurity orbital coupled to $B$ bath orbitals per spin, with $2(B+1)$ spin-orbitals. Its cost can therefore grow exponentially with $B$. In addition, the equilibrium starting point can become numerically fragile for large $B$. Applications so far reached $B=7$~\cite{Guerci2023TDgGA}. This already captures the relaxation qualitatively. For the quenches studied here, it is converged in $B$ to an absolute error of $10^{-3}$ in the double occupancy up to $t\simeq 4$--$5$. The gGA variational landscape is degenerate at $U=0$ for $B\geq 3$. These calculations therefore started from the equilibrium state at a small but finite initial interaction, $U_i=0.05$, rather than from the Fermi sea used in DMFT.

\begin{figure}[t]
	\centering
	\includegraphics[width=\columnwidth]{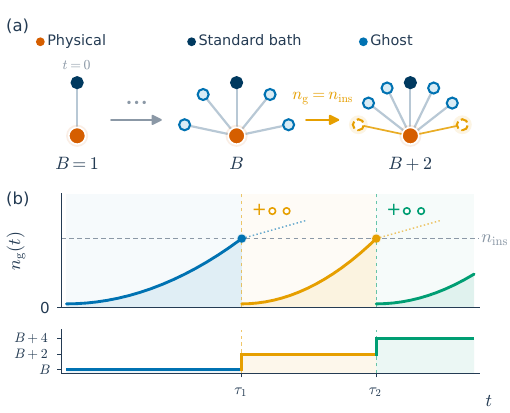}
	\caption{
	\textbf{Adaptive td-gGA.} (a) In the embedding problem of each site, the physical orbital hybridizes with $B$ bath orbitals per spin, shown as lines: one standard bath orbital, as in the Gutzwiller approximation, and $B-1$ ghost orbitals. The dynamics starts from $B=1$, which is exact for the Fermi sea. A new particle-hole-symmetric pair of ghost orbitals, shown dashed, is added when the population $n_\mathrm{g}$ of the outermost pair reaches the threshold $n_\mathrm{ins}$. (b) Schematic time dependence of $n_\mathrm{g}$ and of $B$. Each pair is populated only after a time set by the dynamics, not preset. When $n_\mathrm{g}$ reaches $n_\mathrm{ins}$ at the insertion times $\tau_1$ and $\tau_2$, $B$ grows by two. The population $n_\mathrm{g}$ then follows the new outermost pair, while the previous pair, shown dotted, keeps filling. Insertion stops at $B=B_\mathrm{max}$.
	}
	\label{fig: scheme}
\end{figure}

In this work we introduce adaptive td-gGA, which makes td-gGA a quantitative method for the dynamics of correlated electrons, demonstrated here on this benchmark problem. It takes td-gGA well beyond previous calculations in three respects: from $6$ to $16$ ghost orbitals per spin, from a quench out of a weakly interacting state to the exact Fermi sea, and, at strong coupling, to converged times $t\gtrsim 9$. These times are up to three and a half times as long as those reached by numerically exact DMFT~\cite{Eckstein09ThermalizationIQ}. The scheme is partly motivated by adaptive variational quantum dynamics simulations (AVQDS)~\cite{AVQDS} and their imaginary-time counterpart~\cite{AVQITE}. These construct a compact ansatz adaptively along the real- or imaginary-time path. They add new variational parameters only when the current ansatz can no longer follow the evolution. Here the ghost orbitals play the role of these parameters. The scheme rests on a simple observation: each additional ghost pair is nearly empty at the start and is populated only after a time set by the dynamics itself. The signal is the population $n_\mathrm{g}(t)$ of the outermost ghost pair, that is, the occupation by which its orbitals depart from being empty or filled. We therefore start from $B=1$, which is exact for the Fermi sea. We then insert a new ghost pair whenever $n_\mathrm{g}$ reaches a small threshold $n_\mathrm{ins}$, as sketched in Fig.~\ref{fig: scheme}. Neither an initial interaction nor an equilibrium calculation at large $B$ is required. Adaptive td-gGA and DMFT are thus compared for exactly the same protocol, and a single framework covers weak, intermediate and strong coupling alike. The number of orbitals $B$ is not a preset parameter but is determined by the dynamics itself. The only parameter is its upper bound $B_\mathrm{max}$, set by the available computational resources. Runs with smaller $B_\mathrm{max}$ serve as a convergence check. We write $B\leq B_\mathrm{max}$ for such a run. We establish the following results:
(i) With $B\leq 17$, the dynamics is converged in $B$ up to $t\gtrsim 10$ at all interaction strengths except $U=4$, where it is converged up to $t=9.1$. The double occupancy $d(t)$ and the Fermi-surface discontinuity $\Delta n(t)$ agree with DMFT at weak and strong coupling within the accuracy of the DMFT data, including the prethermal plateau and the revivals. At weak coupling, a small offset of $d(t)$ remains, which lies within the accuracy stated for the DMFT data~\cite{Eckstein09ThermalizationIQ}. At strong coupling, td-gGA reaches times well beyond the DMFT data.
(ii) Near the crossover, the double occupancy relaxes to the thermal gGA value at the effective temperature, which in turn reproduces the DMFT thermal values. At strong coupling, it stays near the prethermal value up to the end of the time window.
(iii) Resolving the dynamics in $B$ shows how the ghost orbitals turn the sharp td-GA transition into the DMFT crossover. It also shows how many ghost orbitals each feature requires.
(iv) At strong coupling and beyond the DMFT time window, the oscillation envelope becomes independent of $U$ at large $U$. Its initial decay is fixed by the width of the band of excitations with one doubly occupied site, a doublon, and one empty site, a holon. The oscillation frequency is renormalized as predicted by a second-order strong-coupling expansion. The oscillation center remains near the prethermal value, consistent with the slow decay of doublons at strong coupling~\cite{Strohmaier2010doublon, Sensarma2010lifetime}.

\section{Model and method}\label{sec: method}

\subsection{Hubbard model and quench protocol}
We consider the single-band Hubbard model on the Bethe lattice with infinite coordination,
\be
\h = \sum_{\braket{I,J}}\sum_{\sigma} t_\mathrm{h}\, \cc_{I\sigma}\ca_{J\sigma} + U \sum_{I} \left(\hat{n}_{I\up}-\tfrac{1}{2}\right)\left(\hat{n}_{I\dw}-\tfrac{1}{2}\right), \label{eq: hubbard}
\ee
where $\cc_{I\sigma}$ creates an electron with spin $\sigma$ on site $I$ and $\hat{n}_{I\sigma}=\cc_{I\sigma}\ca_{I\sigma}$. The sum $\braket{I,J}$ runs over the nearest-neighbor pairs of a Bethe lattice with coordination number $z\to\infty$, in both orders, $IJ$ and $JI$, so that the hopping term is Hermitian. The hopping $t_\mathrm{h}=W/(2\sqrt{z})$ is scaled to keep the half-bandwidth $W$ finite~\cite{ga_infdvollhardt}. We consider half filling, with the semicircular density of states $\rho_0(\eps) = \frac{2}{\pi W}\sqrt{1-(\eps/W)^2}$ of the single-particle energies $\eps$. The half-bandwidth $W=1$ sets the unit of energy, and $\hbar/W$ the unit of time, with $\hbar=1$. The system is prepared in the Fermi sea, and the interaction is switched on suddenly to $U$ at $t=0$. Energy is conserved, $E = E_0 = -4/(3\pi)$, the kinetic energy of the Fermi sea. We monitor the double occupancy $d(t)=\braket{\hat{n}_{I\up}\hat{n}_{I\dw}}$ and the momentum distribution $n(\eps,t)$, the occupation of the single-particle states of energy $\eps$. Of the latter we focus on its discontinuity at the Fermi level, $\Delta n(t) = n(0^-,t)-n(0^+,t)$. The DMFT results of Ref.~\cite{Eckstein09ThermalizationIQ} are quoted in units of the quarter-bandwidth $V$ and time $1/V$. With $U_\mathrm{EKW}$ and $t_\mathrm{EKW}$ the interaction and time in those units, $U = U_\mathrm{EKW}/2$ and $t = 2t_\mathrm{EKW}$ in ours. Reference~\cite{Eckstein09ThermalizationIQ} brackets the DMFT crossover between $3V$ and $3.3V$ and estimates it from its data at $3.2V$. In our units this is $U_c^\mathrm{DMFT}=1.6$, within $1.5\leq U\leq 1.65$.

The exact short-time curvature and the leading strong-coupling result serve as checks and references. At short times, energy conservation at half filling, $E_\mathrm{kin}(t) + U[d(t)-\tfrac{1}{4}] = E_0$ per site, gives $\ddot{d} = -\ddot{E}_\mathrm{kin}/U$. The Fermi sea is an eigenstate of the hopping term $\hat{T}$ of Eq.~\eqref{eq: hubbard}. The second derivative at $t=0$ therefore reduces to $\ddot{E}_\mathrm{kin}(0) = -U^2\braket{[[\hat{T},\hat{D}],\hat{D}]}/L$ for a lattice of $L$ sites, where $\hat{D}=\sum_I\hat{n}_{I\up}\hat{n}_{I\dw}$ counts the doubly occupied sites. This operator differs from the interaction term of Eq.~\eqref{eq: hubbard} divided by $U$ only by a function of the conserved particle number. The double commutator multiplies each hopping term $\cc_{I\sigma}\ca_{J\sigma}$ by $(\hat{n}_{I\sbar}-\hat{n}_{J\sbar})^2$, where $\sbar$ is the spin opposite to $\sigma$. In the Fermi sea the two spin species are uncorrelated. Hence $\braket{(\hat{n}_{I\sbar}-\hat{n}_{J\sbar})^2} = \tfrac{1}{2}$ up to corrections of order $1/z$, which vanish for $z\to\infty$. It gives $\ddot{d}(0) = U E_0/2$. Since the Hamiltonian and the Fermi sea are real, $d(t)$ is even in $t$, so that
\be
d(t) = \frac{1}{4} - \frac{U t^2}{3\pi} + \bigO(t^4), \label{eq: short}
\ee
and at strong coupling, to first order in the hopping and at times short compared with the inverse width of the doublon-holon band,
\be
d^{(1)}(t) = \frac{1}{4} - \frac{|E_0|}{2U}(1-\cos Ut). \label{eq: d1}
\ee
Equation~\eqref{eq: d1} is the first-order result of Ref.~\cite{Eckstein09ThermalizationIQ} with its envelope at the initial value. It oscillates with period $2\pi/U$ around the prethermal value $d_\mathrm{stat} = \frac{1}{4}-\frac{|E_0|}{2U}$. At later times, dephasing within the doublon-holon band damps the oscillation by an envelope, discussed in Sec.~\ref{sec: strong}, while its center stays at $d_\mathrm{stat}$. The high-temperature thermal value, in contrast, is $d^{(1)}_\mathrm{th}=\frac{1}{4}-\frac{|E_0|}{U}$, so that $d(t)$ relaxes only halfway to it~\cite{Eckstein09ThermalizationIQ}. Equation~\eqref{eq: short} completes the digitized DMFT curves near $t=0$ in Sec.~\ref{sec: convergence} and tests td-gGA at short times. Equation~\eqref{eq: d1}, $d_\mathrm{stat}$ and $d^{(1)}_\mathrm{th}$ set the reference for the strong-coupling dynamics in Sec.~\ref{sec: strong}, and Appendix~\ref{app: c} derives Eq.~\eqref{eq: d1} from the strong-coupling expansion.

\subsection{Time-dependent ghost Gutzwiller approximation}\label{sec: tdgga}
In td-gGA~\cite{Guerci2023TDgGA} the many-body state is $\ket{\Psi_\mathrm{G}(t)} = \hat{\mathcal{P}}_\mathrm{G}(t)\ket{\Psi_0(t)}$. Here $\ket{\Psi_0(t)}$ is a Slater determinant of quasiparticles in an enlarged space of $B$ orbitals per spin and site: one standard bath orbital, as in the Gutzwiller approximation, and $B-1$ ghost orbitals. They become the bath orbitals of the embedding problem. The operator $\hat{\mathcal{P}}_\mathrm{G}(t)$ is a product of local maps to the physical Fock space. For a single band with density of states $\rho_0$, as on the Bethe lattice, the equations of motion reduce to two objects. The first is a quasiparticle density matrix $\nu_{ab}(\eps,t)=\avihi{\Psi_0}{\fc_{\eps a}\fa_{\eps b}}$ on the energy grid of $\rho_0$, where $\fc_{\eps a}$ creates a quasiparticle labeled by the band energy $\eps$ in orbital $a=1,\ldots,B$. The second is a single embedding state $\ket{\Phi(t)}$ of $2(B+1)$ spin-orbitals. They evolve as
\begin{align}
i\partial_t \nu(\eps) &= \eps\left[\nu(\eps),\, (\R\R^\dag)^{\mathrm{T}}\right], \label{eq: eom_n}\\
i\partial_t\ket{\Phi} &= \h^\mathrm{eb}\ket{\Phi}. \label{eq: eom_phi}
\end{align}
The bracket in Eq.~\eqref{eq: eom_n} is the commutator of $B\times B$ matrices. Each $\nu(\eps)$ thus evolves by a unitary rotation and keeps its eigenvalues. The embedding Hamiltonian is
\be
\begin{aligned}
\h^\mathrm{eb} =\; & U\left(\hat{n}_\up-\tfrac{1}{2}\right)\left(\hat{n}_\dw-\tfrac{1}{2}\right) + \sum_{a\sigma}\left(\D_a\cc_\sigma\ba_{a\sigma} + \mathrm{H.c.}\right) \\
& + \sum_{ab\sigma}\lambda^\mathrm{c}_{ab}\ba_{b\sigma}\bc_{a\sigma}.
\end{aligned} \label{eq: heb}
\ee
Here $\cc_\sigma$ and $\hat{n}_\sigma$ refer to the physical orbital of the impurity, and $\bc_{a\sigma}$, $a=1,\ldots,B$, creates an electron in bath orbital $a$. The renormalization vector $\R$, the coupling $\D$ and the bath energies $\lambda^\mathrm{c}$ are fixed at each time by the embedding density matrix $\Delta_{ab}=\avihi{\Phi}{\ba_b\bc_a}$ and by three relations~\cite{Guerci2023TDgGA}: $\avihi{\Phi}{\cc_\sigma\ba_{a\sigma}} = [\R^\mathrm{T}\sqrt{\Delta(1-\Delta)}]_a$, $\int d\eps\,\rho_0(\eps)\,\eps\,[\nu(\eps)\R^*]_a = [\sqrt{\Delta(1-\Delta)}\D]_a$, and $\lambda^\mathrm{c}_{ab} = -\partial\big[\R^\mathrm{T}\sqrt{\Delta(1-\Delta)}\,\D + \mathrm{c.c.}\big]/\partial\Delta_{ab}$. The physical momentum distribution follows from the quasiparticle one,
\be
n(\eps,t) = \R^\mathrm{T} \nu(\eps,t)\,\R^* + \kappa(t),
\ee
where the constant $\kappa$ restores $\int d\eps\, \rho_0(\eps)\, n(\eps,t)=1/2$. The total energy $E = 2\int d\eps\,\rho_0\,\eps\, n(\eps) + U(d-1/4)$ is conserved to an absolute error of $4\times10^{-7}$ or better when the embedding state is propagated exactly. With the low-rank propagation used for $B=17$ below, it is conserved to an absolute error of $1\times10^{-4}$. For $B=1$, $\nu(\eps)$ is static, and the dynamics reduces to a single two-level system~\cite{schiro10tdga}. This system has the dynamical transition at $U_c^\mathrm{dyn}=U_c^\mathrm{BR}/2 = 4|E_0| = 16/(3\pi)\approx 1.698$.

We integrate Eqs.~\eqref{eq: eom_n}--\eqref{eq: eom_phi} with an adaptive Runge-Kutta (RK45) scheme~\cite{Brankin1993rksuite} at relative tolerance $10^{-7}$ on 32 Gauss nodes for the energy integrals. Tightening the tolerance to $10^{-9}$ or doubling the number of Gauss nodes to 64 changes $d(t)$ by less than $10^{-6}$ up to $t=10$, as tested at $U=1.5$ for $B=5$ and $7$. The embedding state of $2(B+1)$ spin-orbitals, $28$ at $B=13$, is propagated exactly, with the Hamiltonian applied in a Kronecker-product form. A right-hand-side evaluation at $B=13$ takes $0.2$~s on 32 cores and $7$~GB of memory. For $B=15$, the embedding state has $32$ spin-orbitals. In the half-filled sector with eight electrons of each spin, it has $\binom{16}{8}^2\approx1.7\times10^{8}$ amplitudes. We propagate the same equations on a graphics processing unit (GPU) with the Dormand-Prince RK45 scheme~\cite{Dormand1980embedded} at the same tolerance. The code exploits the symmetry of the quench state under the exchange of the two spin species. A right-hand-side evaluation takes $0.28$~s on one NVIDIA A100 and $27$~GB of device memory. The resulting $d(t)$ agrees with the CPU code to an absolute difference of $3\times10^{-8}$, up to $t=11.9$ at $U=1.5$ and $t=10.4$ at $U=2.5$. At $B=17$ the embedding state has $\binom{18}{9}^2\approx2.4\times10^{9}$ amplitudes and no longer fits on one GPU. We therefore store it in the low-rank form $\Phi=X\Sigma Y^\dag$, with orthonormal columns of $X$ and $Y$ and a small square matrix $\Sigma$. The factorization is over the single-spin configurations of the two spin species, the cut across which the state is least entangled. We propagate the factors with the rank-adaptive basis-update-and-Galerkin integrator~\cite{Ceruti2022rankadaptive}. Within each time step, $\D$, $\lambda^\mathrm{c}$ and $\R$ are held at the average of their values at the two ends of the step, iterated to self-consistency. The step size is controlled by step doubling, and singular values are discarded up to a weight of $10^{-11}$ per step. At $B=15$ the low-rank propagation reproduces the exact propagation to an absolute difference of $1.3\times10^{-5}$ in $d(t)$, with ranks of $140$--$270$ instead of the $\binom{16}{8}=12870$ configurations per spin. At $B=17$ the rank grows from about $200$ at the insertion of the last pair to $530$--$760$ at $t=10$. Meanwhile the GPU time per unit time grows from $5$ to about $25$ minutes.

\subsection{Adaptive insertion of ghost orbitals}\label{sec: adaptive}
Starting td-gGA from an equilibrium solution at large $B$ can be challenging. For instance, with $B=11$ at the initial interaction $U_i=0.02$, the occupation of the outermost ghost pair is only about $10^{-13}$. This is below the accuracy of the ground-state solver, and the self-consistency loop does not converge. At the same time, a run with fixed $B$ follows the run with $B+2$ until the additional, nearly empty ghost pair fills up. The time at which this happens is not an input but is set by the dynamics itself. A given pair fills at the same time in every run that contains it. Successive pairs fill about $1.5$ time units apart at $U=1.5$ and $1.2$ at $U=2.5$. The ghost pairs thus become relevant one after another.

Adaptive td-gGA exploits this sequential activation of the ghost pairs. It starts with the smallest $B$ that describes the initial state, $B=1$ for the Fermi sea, and adds each ghost pair when the dynamics begins to populate the previous outermost pair, as detected by the criterion below. All td-gGA results in Sec.~\ref{sec: results} are obtained this way. This also removes the need for a finite initial interaction. Starting instead from the gGA equilibrium at small $U_i$ with the same final $U$ shifts the double occupancy by an amount proportional to $U_i$. At $U=1.5$ and $B=5$, the shift is $8\times10^{-4}$ for $U_i=0.01$ and $4\times10^{-3}$ for $U_i=0.05$, already at $t=0$ and up to $t=5$. This is comparable with or larger than the deviations from DMFT discussed in Sec.~\ref{sec: results}. The criterion for adding a pair is the occupation of the outermost ghost pair, which measures the correlations built up by the dynamics. The pair stays empty or filled as long as the inner pairs can describe these correlations. It departs from $0$ or $1$ once they cannot. At every output time, spaced by $0.05$, we therefore monitor its population, $n_\mathrm{g}=\min(\Delta_\mathrm{min}, 1-\Delta_\mathrm{max})$, where $\Delta_\mathrm{min}$ and $\Delta_\mathrm{max}$ are the smallest and largest eigenvalues of the embedding density matrix $\Delta$. When it reaches the threshold $n_\mathrm{ins}=10^{-5}$, the current pair has become active. A new pair, here particle-hole-symmetric and spin-singlet like the quench state, is then seeded with weight $\delta=10^{-8}$ in the embedding state, together with a consistent quasiparticle density matrix, and the evolution continues with $B+2$. For $B=17$, the pair is seeded directly in the low-rank form $\Phi=X\Sigma Y^\dag$ of Sec.~\ref{sec: tdgga}, without forming the dense state. Seeding a vanishing pair leaves all observables unchanged to $\bigO(\delta)$. Adaptive runs reproduce fixed-$B$ runs from equilibrium for $B\leq 9$ to about $10^{-4}$ at the same $U_i$, at a much lower cost. On CPU nodes, adaptive td-gGA from the Fermi sea to $t=10$ costs about $0.08$ core hours for $B\leq 9$ and about $18$ core hours for $B\leq 13$. Fixed-$B$ td-gGA is considerably more expensive. From the gGA equilibrium at $U_i=0.05$, the protocol of Ref.~\cite{Guerci2023TDgGA}, fixed $B=9$ costs about $14$ times more. Most of this cost goes into the equilibrium, which already at $B=11$ does not converge, as noted above. Alternatively, fixed-$B$ runs can start from the Fermi sea with all ghost pairs seeded at $t=0$, which gives the same $d(t)$ to an absolute difference of $3\times10^{-5}$. For the same cost, fixed $B=13$ then covers only about a fifth of the time window: it reaches $t\approx 2$ in the wall time in which adaptive td-gGA reaches $t=10$. On one A100 GPU, adaptive td-gGA with $B\leq 15$ reaches $t=10$ in about $25$ minutes. A fixed $B=15$ run again covers only about a fifth of this time window. The GPU itself contributes a further factor of about $12$ at $B=15$, with $0.28$~s per right-hand-side evaluation against $3.5$~s on a 32-core CPU node.

The scheme works because a new pair is populated by the dynamics of the pairs already present, not by amplifying the seed. The first insertion, at $t=0$ into the Fermi sea, shows this explicitly. To leading order in $t$, the quench turns the physical operator into $\ca_\sigma(t) = \ca_\sigma - \tfrac{i}{2}Ut\,\hat{X}_\sigma$, with the normalized composite operator $\hat{X}_\sigma = 2(\hat{n}_{\bar\sigma}-\tfrac12)\ca_\sigma$. It thereby transfers the spectral weight $U^2t^2/4$ to $\hat{X}_\sigma$. In td-gGA this weight goes to the new pair. We find $|\R_a|^2 = U^2t^2/8$ for each of its two orbitals and $1-|\R_1|^2=U^2t^2/4$ for the original one, to a relative accuracy of $10^{-4}$ for $t\to0$. Hence $\R^\dag\R=1$ to this order. Because $\R$ is defined as a ratio of embedding amplitudes, it is finite however small the occupation of the pair is. Through Eq.~\eqref{eq: eom_n}, the coupling $\eps\,\R_a\R_1^*$ then moves occupied quasiparticle states into the empty ghost orbital. Its occupation, the smallest eigenvalue of $\Delta$, becomes to leading order
\be
\begin{aligned}
n_\mathrm{g}(t) &= \frac{U^2t^4}{32}\left[\int_{-1}^{0}\! d\eps\,\rho_0\,\eps^2 - 2\left(\int_{-1}^{0}\! d\eps\,\rho_0\,\eps\right)^2\right] \\
&= C\,U^2t^4, \qquad C=\frac{1}{256}-\frac{1}{36\pi^2}.
\end{aligned} \label{eq: ng}
\ee
As derived in Appendix~\ref{app: ng}, the second term is the part of the transferred occupation taken up by a rotation of the original orbital. Only the spread of the occupied energies therefore fills the ghost. The constant is $C=1.092\times10^{-3}$. The simulations reproduce it for $t\to0$ to a relative error of $5\times10^{-5}$ at all interaction strengths from $U=0.1$ to $16$, as shown in Appendix~\ref{app: ng}. Equation~\eqref{eq: ng} contains no reference to the seed. Indeed, once the occupation exceeds the seed occupation, it is the same for seed weights from $\delta=10^{-12}$ to $10^{-7}$. The next insertion thus occurs at the same time for all of them. At $U=1.5$, $n_\mathrm{g}$ crosses the threshold at $t=0.25$, and the pair is inserted at the next output time, $t=0.3$. Likewise, the pair added at the next stage reaches the same occupation at a given time whether it is inserted at $t=0.85$ or at $t=2.75$. The seed therefore only fixes the orbitals of the new pair. It also regularizes the factor $[\Delta(1-\Delta)]^{-1/2}$ in the equations of motion, which is singular for an exactly empty pair. Once the driven occupation exceeds the seed, the dynamics retains no memory of it, and the limit $\delta\to0$ is regular. At $U=1.5$ and $B\leq5$, seeds $\delta=10^{-12}$, $10^{-10}$ and $10^{-8}$ give the same $d(t)$ to an absolute difference of $3\times10^{-5}$. At $B\leq7$, random seed directions change it by less than $2\times10^{-5}$. A fixed-$B$ run from equilibrium follows the same mechanism, with the equilibrium occupation of the outer pairs playing the role of the seed. At $U=1.5$ and $B=5$, the two ghost pairs of a run from the equilibrium at $U_i=0.02$ start at $8\times10^{-7}$ and $1\times10^{-8}$. They join the occupations of the adaptive run within $2\%$ and $25\%$ by $t=1.1$ and within $3\%$ by $t=1.5$. The remaining difference in $d(t)$ is that of the initial states at the same $U_i$, the $B=5$ against the $B=1$ equilibrium, $1\times10^{-4}$ per $0.01$ in $U_i$ at $t=0$, which vanishes for $U_i\to0$. The only requirement is to insert a pair before it should have been populated. A late insertion misses this occupation. At $U=1.5$ and $B\leq7$, thresholds from $10^{-7}$ to $10^{-5}$ give the same $d(t)$ to an absolute difference of $2\times10^{-5}$, whereas $10^{-4}$ and $10^{-3}$ shift it by $1\times10^{-4}$ and $4\times10^{-4}$. A seed that is too large biases the result in the same way, by $3\times10^{-4}$ for $\delta=10^{-6}$.

Because the last allowed pair eventually fills too, a run with $B\leq B_\mathrm{max}$ is reliable only up to a convergence time $t_\mathrm{conv}$. It can be estimated from the run itself, without a run with $B\leq B_\mathrm{max}+2$. The outermost pair of the last stage reaches the insertion threshold at a time $t_\mathrm{ins}$, at which a larger run would insert its next pair. We use a tolerance of $10^{-3}$ in $d(t)$, the accuracy of the digitized DMFT double occupancy of Sec.~\ref{sec: convergence}. Runs with $B\leq B_\mathrm{max}$ and $B\leq B_\mathrm{max}+2$ depart from each other by this amount a nearly constant time after $t_\mathrm{ins}$. This lag is $2.2$--$3.8$ for $1.5\leq U\leq 4$ and longer at larger $U$. It is the same for the pairs inserted into $B=11$ and $B=13$ to within $0.3$ time units. We therefore define $t_\mathrm{conv}$ of the $B\leq 15$ runs as $t_\mathrm{ins}$ plus the lag between $B\leq 13$ and $B\leq 11$ at the same $U$. Where both can be evaluated, this agrees to within $0.3$ with the time at which $B\leq 15$ departs from $B\leq 13$ by $10^{-3}$, plus one pair spacing. For $2.5\leq U\leq 4$, $B\leq 15$ is converged only up to $t=7.9$--$9.9$. There we continue with $B\leq 17$, whose $t_\mathrm{conv}$ is the time at which it departs from $B\leq 15$ by $10^{-3}$, plus one pair spacing.

\subsection{Thermal reference}
If the quenched system thermalizes, local observables approach their values in the Gibbs state at the effective temperature $T_\mathrm{eff}$ fixed by $E(T_\mathrm{eff},U) = E_0$~\cite{Eckstein09ThermalizationIQ}. We compute this end point with the finite-temperature extension of gGA~\cite{finiteTgGA2026} as implemented in GEM~\cite{GEM2026}. At the relevant temperatures, the thermal results are converged in $B$ already at $B=3$, with $|d_{B=3}-d_{B=5}|\leq 6\times10^{-5}$. We denote the thermal double occupancy by $d_\infty$.

\section{Results}\label{sec: results}

\begin{figure*}[t]
	\centering
	\includegraphics[width=\textwidth]{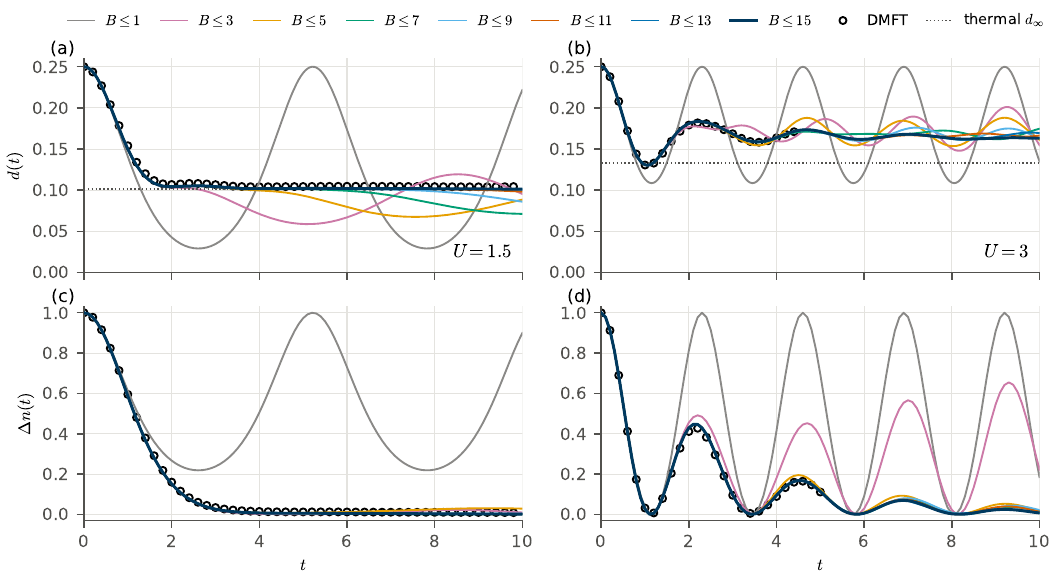}
	\caption{
	\textbf{Convergence in the number of ghost orbitals.} Double occupancy $d(t)$ (a, b) and Fermi-surface discontinuity $\Delta n(t)$ (c, d) after quenches to $U=1.5$ in the left and $U=3$ in the right panels, with adaptive ghost insertion, $B\leq 1, 3, \ldots, 15$. The results are compared with nonequilibrium DMFT~\cite{Eckstein09ThermalizationIQ}, shown as open circles, and with the thermal gGA value $d_\infty$, shown as a dotted line. The case $B\leq 1$ is the conventional td-GA.
	}
	\label{fig: convergence}
\end{figure*}

\subsection{Convergence in \texorpdfstring{$B$}{B}}\label{sec: convergence}
Figure~\ref{fig: convergence} shows the hierarchy $B\leq 1,3,\ldots,15$ for a quench to the weak side of the crossover, $U=1.5$, and to the collapse-and-revival side, $U=3$. The DMFT curves of Ref.~\cite{Eckstein09ThermalizationIQ} were digitized from the published figure, with the axes calibrated by its tick marks. The scatter of the raw points is about $1\times10^{-3}$ for $d(t)$ and $3$--$8\times10^{-3}$ for $\Delta n(t)$, which we take as the accuracy of the digitized curves. For $d(t)$ this is below the accuracy of the DMFT data themselves. Reference~\cite{Eckstein09ThermalizationIQ} estimated that accuracy from the drift of the conserved total energy and stated it to be of the order of the symbol size in the published figure, which corresponds to about $4\times10^{-3}$ in $d$. Near $t=0$ the published curves overlap. There we completed them with the exact expansion~\eqref{eq: short}, which td-gGA reproduces at all $U$, with a $t^2$ coefficient equal to $U/(3\pi)$ to a relative accuracy of $10^{-4}$. The td-GA, $B=1$, oscillates without damping between $d=1/4$ and a minimum. Its Fermi-surface discontinuity revives to $\Delta n=1$ at every period. Each additional ghost pair extends the time window in which td-gGA follows DMFT, by roughly $1.5$ time units. The runs depart from each other in the order in which the ghost pairs fill. With $B\leq 13$ and $15$, the double occupancy follows DMFT up to $t=10$ at $U=1.5$ and up to the end of the DMFT data at $t=4.5$ for $U=3$. At $U=3$ the maximum absolute deviation is $2.3\times10^{-3}$, which $B\leq 7$ already reaches. The momentum distribution converges faster than $d(t)$: $\Delta n(t)$ is converged for $B\leq 5$ up to $t\approx 6$ at $U=1.5$. At $U=3$ the first revival is converged for $B\leq 5$, which differs from $B\leq 15$ by at most $4\times10^{-4}$ in $\Delta n$ there. The converged maximum is $0.449$ at $t=2.15$, against $0.430$ at $t=2.15$ for DMFT. The second revival is converged for $B\leq 7$, with $0.169$ at $t=4.50$ against $0.166$ at $t=4.55$ for DMFT. In contrast, $B\leq 3$ overestimates the second revival almost threefold.

\begin{figure*}[t]
	\centering
	\includegraphics[width=\textwidth]{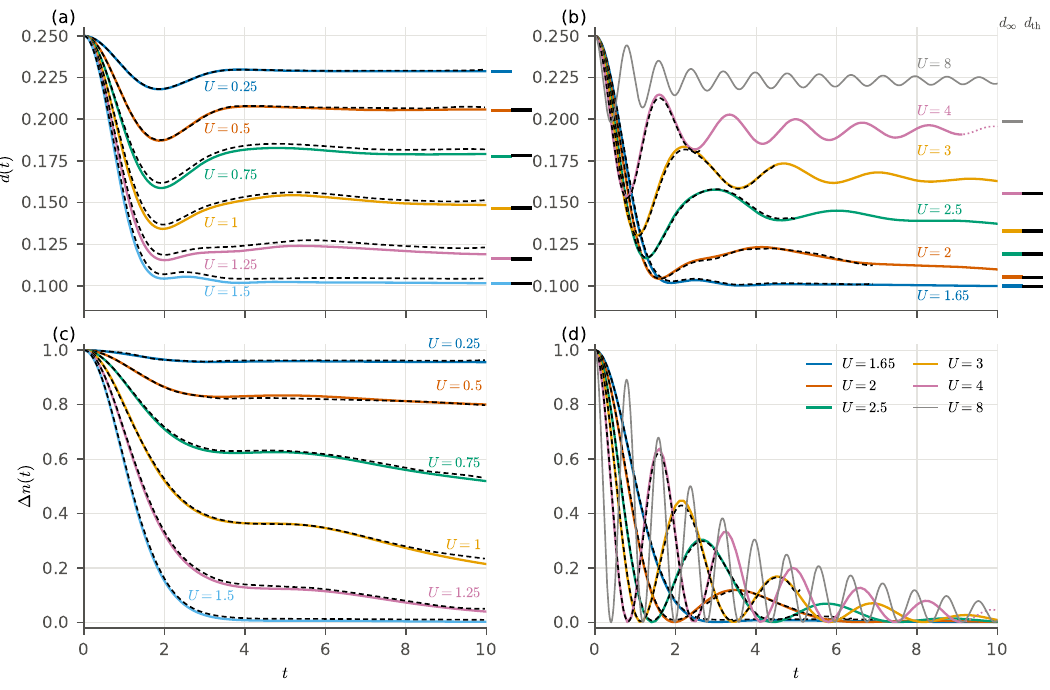}
	\caption{
	\textbf{td-gGA versus DMFT at all interaction strengths.} Double occupancy (a, b) and Fermi-surface discontinuity (c, d) for quenches from $U=0$ to the values of Ref.~\cite{Eckstein09ThermalizationIQ}, at weak coupling (a, c) and strong coupling (b, d). The case $U=8$, shown as a thin gray line, lies beyond the parameter range covered by Ref.~\cite{Eckstein09ThermalizationIQ}. Solid lines show td-gGA with $B\leq 15$, and $B\leq 17$ for $2.5\leq U\leq 4$, up to $t_\mathrm{conv}$ of Table~\ref{tab: dmft}, and dotted beyond. Dashed lines show DMFT~\cite{Eckstein09ThermalizationIQ}. The bars at the right edge of (a, b) are the thermal values, $d_\infty$ from thermal gGA in color and $d_\mathrm{th}$ from DMFT in black.
	}
	\label{fig: dmft}
\end{figure*}

\begin{table}[!tb]
\caption{Convergence, thermal end point and agreement with DMFT for the quenches $0\to U$ from the Fermi sea. Listed are the convergence time $t_\mathrm{conv}$ of the $B\leq 15$ run, $B\leq 17$ for $2.5\leq U\leq 4$, the effective temperature $T_\mathrm{eff}$, the thermal double occupancy $d_\infty$ from thermal gGA and $d_\mathrm{th}$ from DMFT~\cite{Eckstein09ThermalizationIQ}, and the root-mean-square (rms) deviations of td-gGA from DMFT for $d(t)$ and $\Delta n(t)$ over $0.2\leq t\leq\min(t_\mathrm{conv}, t_\mathrm{DMFT})$. The runs end at $t=12$ for $B\leq 15$ and at $t=10$ for $B\leq 17$. Where $t_\mathrm{conv}$ exceeds this time, it is given as $>12$ or $>10$. The window leaves out the first $0.2$ time units, where both methods follow the exact expansion~\eqref{eq: short} and would only dilute the rms. The DMFT data end at $t_\mathrm{DMFT}=10$ for $U\leq 1.5$, $6.9$ for $U=1.65$ and $2$, and $5.0$, $4.5$ and $2.6$ for $U=2.5$, $3$ and $4$. The values of $d_\mathrm{th}$ are read from the arrows in Fig.~2 of Ref.~\cite{Eckstein09ThermalizationIQ} and are accurate to about $1\times10^{-3}$.}
\label{tab: dmft}
\begin{ruledtabular}
\begin{tabular}{ccccccc}
$U$ & $t_\mathrm{conv}$ & $T_\mathrm{eff}$ & $d_\infty$ & $d_\mathrm{th}$ & rms $d$ & rms $\Delta n$ \\
\hline
0.25 & $>12$ & 0.035 & 0.2288 & --     & 0.0003 & 0.004 \\
0.5  & $>12$ & 0.069 & 0.2052 & 0.2055 & 0.0008 & 0.006 \\
0.75 & $>12$ & 0.104 & 0.1778 & 0.1782 & 0.0026 & 0.007 \\
1    & $>12$ & 0.147 & 0.1464 & 0.1467 & 0.0021 & 0.006 \\
1.25 & $>12$ & 0.220 & 0.1162 & 0.1161 & 0.0031 & 0.008 \\
1.5  & $>12$ & 0.338 & 0.1014 & 0.1012 & 0.0026 & 0.007 \\
1.65 & 11.3 & 0.422 & 0.0999 & 0.0996 & 0.0009 & 0.005 \\
2    & 10.1 & 0.638 & 0.1053 & 0.1050 & 0.0008 & 0.006 \\
2.5  & 10.9 & 0.997 & 0.1192 & 0.1190 & 0.0009 & 0.005 \\
3    & $>10$ & 1.42  & 0.1329 & 0.1328 & 0.0011 & 0.008 \\
4    & 9.1  & 2.48  & 0.1552 & 0.1552 & 0.0014 & 0.008 \\
\end{tabular}
\end{ruledtabular}
\end{table}

\subsection{Comparison with DMFT at all interaction strengths}\label{sec: dmft}
Figure~\ref{fig: dmft} and Table~\ref{tab: dmft} compare td-gGA with $B\leq 15$, and $B\leq 17$ for $2.5\leq U\leq 4$, with DMFT at all eleven quenches of Ref.~\cite{Eckstein09ThermalizationIQ}. As listed in Table~\ref{tab: dmft}, $t_\mathrm{conv}$ lies beyond $t=12$ for $U\leq 1.5$, at $11.3$ for $U=1.65$, at $9.1$--$10.9$ for $U=2$, $2.5$ and $4$, and beyond $10$ for $U=3$. The dynamics is thus converged in $B$ up to $t\gtrsim 10$, the end of the window of Figs.~\ref{fig: convergence} and \ref{fig: dmft}, at all interaction strengths except $U=4$, where $t_\mathrm{conv}=9.1$. At a tolerance of $4\times10^{-3}$, the accuracy stated for the DMFT data~\cite{Eckstein09ThermalizationIQ}, $t_\mathrm{conv}$ exceeds $t=10$ also at $U=4$, since $B\leq 17$ and $B\leq 15$ differ there by at most $1.1\times10^{-3}$ up to $t=10$. The Fermi-surface discontinuity agrees with DMFT at all $U$ to an rms of $0.004$--$0.008$, comparable with the digitization accuracy. The largest deviations, up to $0.02$, occur on the steep parts of the curves, where a small shift in time dominates.

On the strong-coupling side, $U\geq 1.65$, the double occupancy agrees with DMFT to an rms of $0.8$--$1.4\times10^{-3}$ and a maximum absolute deviation of $2.3\times10^{-3}$. The rms is comparable with the accuracy of the digitized curves, and the maximum deviation lies within the accuracy of the DMFT data. The extrema agree to $0.01$--$0.06$ in time. At $U=4$, for instance, td-gGA puts the first minimum at $t=0.789$ and the first maximum at $1.596$, and DMFT at $0.793$ and $1.598$. The collapse and revival of the Fermi-surface discontinuity is reproduced in position and height. The first revival of $\Delta n(t)$ reaches the height $\Delta n_1 = 0.304$ at $t=2.60$ for $U=2.5$, $0.449$ at $2.15$ for $U=3$ and $0.638$ at $1.60$ for $U=4$. DMFT gives $0.299$ at $2.66$, $0.430$ at $2.15$ and $0.622$ at $1.58$. At $U=3$ the second revival agrees as well, $0.169$ at $t=4.50$ against $0.166$ at $4.55$.

On the weak-coupling side, td-gGA reproduces the rapid drop of $d(t)$ and the subsequent shallow oscillation. It also reproduces the prethermal plateau of $\Delta n(t)$~\cite{Moeckel08InteractionQH} together with its slow decay toward zero. At $U=0.75$, for instance, $\Delta n(10) = 0.520$ against $0.530$ in DMFT. From $t\approx 2$ on, however, $d(t)$ lies systematically below DMFT. The offset is $0.002$--$0.003$ for $U=0.75$--$1.5$, with a maximum deviation of $4\times10^{-3}$, and below $0.001$ for $U\leq 0.5$. It is the same for every $B$ once the run is converged. The difference is within the accuracy of the DMFT double occupancy, about $4\times10^{-3}$. The thermal values provide an independent reference. At $U=1.5$, close to the crossover, both methods find fast thermalization. There the mean of the td-gGA $d(t)$ over $8\leq t\leq 10$ is $0.1015$, at the thermal value $d_\infty=0.1014$. The DMFT curve at $t=9.4$--$9.8$ lies at $0.104$, slightly above the DMFT thermal value $d_\mathrm{th}=0.1012$. This thermal value agrees with thermal gGA to an absolute difference of $2\times10^{-4}$. The weak-coupling offset is thus within the combined accuracy of the two calculations.

The thermal end point is accurate at all $U$. Thermal gGA at $T_\mathrm{eff}$ reproduces the DMFT thermal double occupancy $d_\mathrm{th}$ to within an absolute difference of $4\times10^{-4}$ at all ten interaction strengths where it is available, as listed in Table~\ref{tab: dmft}. At $U=1.5$, td-gGA reaches this thermal value within the simulated time, as shown above. A sharper test is the full momentum distribution. Let $n_\infty(\eps)$ be the thermal gGA distribution at $T_\mathrm{eff}=0.338$. At $U=1.5$, $\int d\eps\,\rho_0(\eps)\,|n(\eps,t) - n_\infty(\eps)|$ falls from $0.32$ at $t=0$ to $6\times 10^{-3}$ at $t=3$ and $5\times10^{-4}$ at $t=10$. The remaining deviation consists of a small residual discontinuity at $\eps=0$ and a smooth odd profile of a few $10^{-3}$. This profile decays roughly exponentially on a time scale of about $2.5$. Within the converged window, the physical one-body distribution of td-gGA thus approaches the thermal one, not only the double occupancy. The td-gGA state remains pure, and Eq.~\eqref{eq: eom_n} conserves the spectrum of $\nu(\eps)$. This relaxation can therefore happen only at the level of the physical observables, with the growing set of ghost orbitals acting as an effective purifying bath~\cite{Verstraete2004MatrixPD}.

\begin{figure*}[t]
	\centering
	\includegraphics[width=\textwidth]{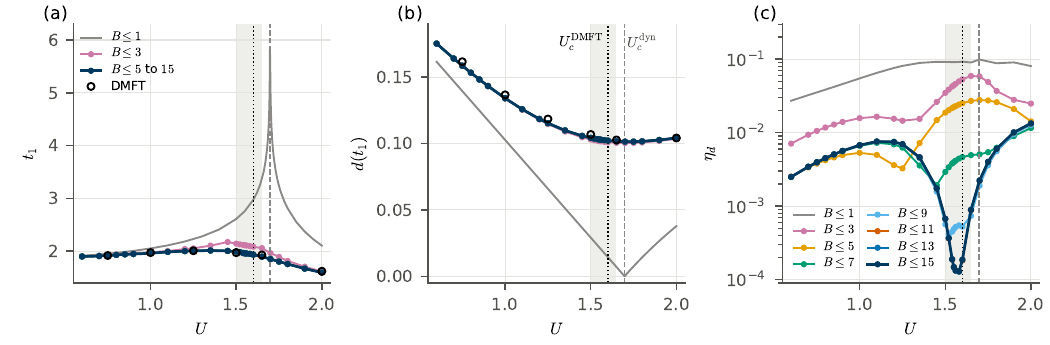}
	\caption{
	\textbf{From the td-GA transition to the dynamical crossover.} (a) Time $t_1$ of the first minimum of the double occupancy, half the oscillation period in td-GA, and (b) the double occupancy $d(t_1)$ at this minimum versus $U$ for $B\leq 1, \ldots, 15$. DMFT~\cite{Eckstein09ThermalizationIQ} is shown as open circles. For $B\leq 3$ near the crossover, where $d(t)$ has only a shoulder, $t_1$ is the first stationary point. The td-GA curve, $B=1$, is computed on a finer grid in $U$ that resolves the divergence at $U_c^\mathrm{dyn}$. (c) Root-mean-square distance $\eta_d$ of $d(t)$ from the thermal value $d_\infty$ over $4\leq t\leq 7$. The curves for $B\leq 5$ to $15$ nearly coincide in (a) and (b), and those for $B\leq 11$ to $15$ in (c). The shaded band is the range $1.5\leq U\leq 1.65$ to which the DMFT data restrict the crossover~\cite{Eckstein09ThermalizationIQ}. The vertical lines mark the DMFT estimate of the crossover, $U=1.6$, dotted, and the td-GA transition $U_c^\mathrm{dyn}=16/(3\pi)$, dashed.
	}
	\label{fig: crossover}
\end{figure*}

\subsection{From the td-GA transition to the dynamical crossover}
To see how ghost orbitals resolve the sharp td-GA transition, we repeated the quench for 25 interaction strengths between $U=0.6$ and $2$, with $B\leq 1, 3, \ldots, 15$ each. The step in $U$ is $0.02$ around the crossover. Up to $t=7$, the $B\leq 15$ runs agree with $B\leq 13$ to an absolute difference of $4\times10^{-6}$ in $d(t)$ at all $U$. Two diagnostics characterize the result. The first follows the td-GA transition itself. In td-GA, $d(t)$ oscillates between $1/4$ and its first minimum $d(t_1)$, so that the time $t_1$ of the first minimum is half the oscillation period. With ghost orbitals, we take $t_1$ as the first stationary point of $d(t)$, the first time at which $\dot{d}$ vanishes or, where $d(t)$ has only a shoulder, comes closest to zero. This is the first minimum of $d(t)$, except for $B\leq 3$ near the crossover. Figures~\figref{fig: crossover}{(a) and (b)} show $t_1$ and $d(t_1)$. The second diagnostic measures relaxation. It is the rms distance of $d(t)$ from $d_\infty$ over $4\leq t\leq 7$,
\begin{equation}
\eta_d=\Big\{\frac13\int_4^7 dt\,[d(t)-d_\infty]^2\Big\}^{1/2},
\end{equation}
shown in Fig.~\figref{fig: crossover}{(c)}. The window starts at $t=4$, after the initial drop and the first oscillation of $d(t)$, and ends at $t=7$, where the runs with $B\leq 11$ to $15$ still agree, so that $\eta_d$ is converged in $B$.

In td-GA, both $t_1$ and $d(t_1)$ are singular at $U_c^\mathrm{dyn}=1.698$. As $U$ approaches it from either side, the half-period diverges logarithmically, $t_1\simeq -(3\pi/16)\ln|U-U_c^\mathrm{dyn}|$~\cite{schiro10tdga}, from $2.6$ at $U=1.5$ to $5.9$ at $U=1.697$. Our td-GA data follow this law with slopes of $-0.58$ and $-0.60$ below and above $U_c^\mathrm{dyn}$. At the same time, the minimum $d(t_1)$ decreases linearly to zero. At the transition the doublons are thus depleted completely. Ghost orbitals remove this singularity. Already for $B\leq 3$, $t_1$ and $d(t_1)$ are smooth functions of $U$. They differ from the $B\leq 15$ values by at most $0.17$ and $1.3\times10^{-3}$. The largest differences occur in the crossover, where $d(t)$ for $B\leq 3$ has only a shoulder. From $B\leq 5$ on, the maximum absolute differences are $10^{-3}$ and $2\times10^{-6}$. The converged values agree with DMFT at the six interactions of Ref.~\cite{Eckstein09ThermalizationIQ} to within $0.03$ in $t_1$ and $0.003$ in $d(t_1)$. Instead of the singularities of td-GA, $t_1$ has a broad maximum of $2.0$ and $d(t_1)$ a shallow minimum of $0.10$. Through the crossover, $U=1.5$--$1.65$, they vary only between $1.90$ and $2.00$ and between $0.102$ and $0.104$.

The relaxation of $d(t)$ requires more ghost orbitals and locates the crossover. In td-GA, $d(t)$ oscillates without damping at every $U$ and stays $0.03$--$0.1$ away from $d_\infty$. For $B\leq 3$ and $5$, $\eta_d$ has a shallow minimum at $U=1.25$ and a pronounced maximum at $U=1.65$--$1.7$, close to $U_c^\mathrm{dyn}$. From $B\leq 7$ on, the maximum is gone. The minimum deepens and moves with $B$, from $U=1.45$ for $B\leq 7$ to $1.52$ for $B\leq 9$ and $1.58$ for $B\leq 11$ to $15$. From $B\leq 11$ on, the minimum is sharp. There $\eta_d=1.3\times10^{-4}$ at $U=1.56$--$1.58$, against $2.5\times10^{-3}$ at $U=0.6$ and $1.3\times10^{-2}$ at $U=2$. The fastest thermalization of td-gGA thus occurs at $U\simeq 1.55$--$1.6$, and locating it requires $B\geq 11$. It lies inside the DMFT bracket of the crossover, $U=1.5$--$1.65$, and close to the DMFT estimate $U=1.6$~\cite{Eckstein09ThermalizationIQ}. Since our grid in $U$ is finer than that of the DMFT data, the precision of this comparison is set by the DMFT bracket. The td-GA transition, only $0.1$ above this estimate, is thus the $B=1$ counterpart of the crossover. The ghost orbitals remove its singularity and place the fastest thermalization inside the DMFT bracket.

\begin{figure*}[t]
	\centering
	\includegraphics[width=\textwidth]{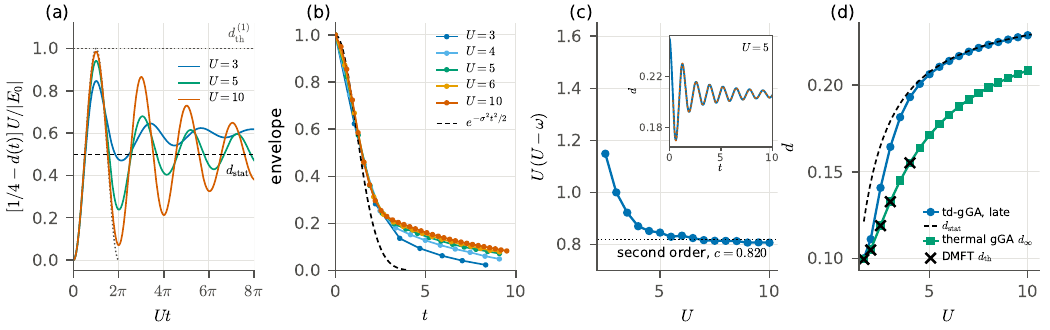}
	\caption{
	\textbf{Strong coupling beyond the DMFT time window.} (a) Double occupancy of td-gGA for $U=3$, $5$ and $10$, scaled as $[1/4-d(t)]\,U/|E_0|$ and plotted against $Ut$ over four periods, with $B\leq 17$. On this scale the first-order result of Eq.~\eqref{eq: d1} is $(1-\cos Ut)/2$, shown dotted over its first period. It oscillates around $d_\mathrm{stat}$ at $0.5$, and $d^{(1)}_\mathrm{th}$ corresponds to $1$. (b) Envelope of the oscillation for $U=3$, $4$, $5$, $6$ and $10$. Each marker is an extremum of the scaled double occupancy of (a), measured from the straight line through its two neighboring extrema. The dashed line is the short-time result $e^{-\sigma^2t^2/2}$ with $\sigma^2=3/4$. (c) Renormalization of the oscillation frequency, $U(U-\omega)$, for $U=2.5$ to $10$ in steps of $0.5$. Here $\omega$ is the angular frequency of damped-cosine fits of $d(t)$ with the envelope of Eq.~\eqref{eq: env}. The inset compares $d(t)$, solid, with its fit, dashed, for $U=5$. The second-order strong-coupling coefficient $c=0.820$ of Appendix~\ref{app: c} is shown dotted. (d) Late-time level of $d(t)$, the mean over the last two periods $2\pi/U$ before $t=10$, which averages out the oscillation. It is compared with the prethermal value $d_\mathrm{stat}$, dashed, the thermal gGA value $d_\infty$ and the DMFT thermal values $d_\mathrm{th}$~\cite{Eckstein09ThermalizationIQ}.
	}
	\label{fig: strong}
\end{figure*}

\subsection{Strong coupling at extended times}\label{sec: strong}
The quench dynamics at weak coupling is well established numerically. The real-time weak-coupling quantum Monte Carlo solution of Ref.~\cite{Eckstein09ThermalizationIQ} reaches $t=10$ for $U\leq 1.5$, and td-gGA agrees with it there, as shown in Sec.~\ref{sec: dmft}. From intermediate to strong coupling, the numerical picture is incomplete. The sign problem of the same solver grows with the interaction and with the simulated time. For $U\geq 2$ the DMFT data therefore cover about two oscillation periods or fewer, and $U>4$ lies beyond the parameter range covered by Ref.~\cite{Eckstein09ThermalizationIQ}. The first-order strong-coupling expansion of Ref.~\cite{Eckstein09ThermalizationIQ} contains the damping of the oscillation only through an envelope that it does not evaluate, and it describes neither a renormalization of the frequency nor a relaxation of the center. Settling these questions requires dynamics that is converged over many oscillation periods throughout the intermediate- and strong-coupling regime. With $B\leq 15$ at $U=2$ and $B\leq 17$ for $2.5\leq U\leq 4$, td-gGA is converged up to $t\geq 9.1$, the lowest at $U=4$. We extend it with $B\leq 17$ to $U=10$ in steps of $0.5$. Above $U=4$, $B\leq 17$ agrees with $B\leq 15$ to within an absolute difference of $4\times10^{-4}$ in $d(t)$ up to $t=10$, so the dynamics is converged over the whole window $t\leq 10$.

We first compare the first oscillation period with the strong-coupling limit. Figure~\figref{fig: strong}{(a)} shows $d(t)$ against $Ut$, scaled such that the first-order result $d^{(1)}(t)$ of Eq.~\eqref{eq: d1} becomes $(1-\cos Ut)/2$ for every $U$. Over the first period, the td-gGA curves approach this result as $U$ grows. Their first peak, at $t=\pi/U$, reaches $0.846$, $0.942$ and $0.985$ for $U=3$, $5$ and $10$, against $1$ at first order. The difference from $1$ decreases as about $1.5/U^2$, that is, at second order in the hopping. Over the whole first period, $0\leq t\leq 2\pi/U$, the absolute rms deviation of $d(t)$ from $d^{(1)}(t)$ decreases from $4\times10^{-2}$ at $U=2.5$ to $1\times10^{-3}$ at $U=10$. Beyond the first period, the extended time window reveals three features, shown in Figs.~\figref{fig: strong}{(b)--(d)}. The envelope becomes a universal function of time, and the oscillation frequency is renormalized. The center of the oscillation stays near the prethermal value of the strong-coupling expansion instead of relaxing to the thermal one~\cite{Eckstein09ThermalizationIQ}.

The first feature is that the oscillation envelope becomes independent of $U$. We measure it at each extremum of $[1/4-d(t)]\,U/|E_0|$, as the height of the extremum above or below the straight line through its two neighboring extrema. This removes a drifting center. In Eq.~\eqref{eq: d1}, where the envelope is held at its initial value, it is $1$ at all times. For $U\geq 6$, the envelopes differ from the one at $U=10$ by at most $0.015$ in absolute terms over the whole window, as shown in Fig.~\figref{fig: strong}{(b)}. For smaller $U$ they lie below it, with maximum absolute differences of $0.08$, $0.04$ and $0.03$ for $U=3$, $4$ and $5$, and approach it as $U$ grows. This common envelope is $|F_0(t)/F_0(0)|$ of the first-order strong-coupling result of Ref.~\cite{Eckstein09ThermalizationIQ}, with $F_0(t)=\braket{e^{i\hat{T}_0t}\hat{T}_+e^{-i\hat{T}_0t}}_0$. Here $\hat{T}_+$ is the part of the hopping that creates a doublon-holon pair, $\hat{T}_0$ the part that conserves the number of doublons, and $\braket{\cdot}_0$ the expectation value in the Fermi sea. That work shows that this envelope decays by dephasing within the doublon-holon band, but does not evaluate it. Its decay at short times is fixed by the width of that band. The expansion of $F_0(t)$ in Appendix~\ref{app: c} gives $|F_0(t)/F_0(0)|=1-\sigma^2t^2/2+\bigO(t^4)$, where $\sigma^2=3/4$ on the Bethe lattice is the variance of the doublon-holon band. At $U=10$, the td-gGA envelope follows $e^{-\sigma^2t^2/2}$ to within $0.02$ up to $t\simeq 1$, without an adjustable parameter. Once the envelope has fallen to about a quarter near $t=2.5$, its decay becomes much slower. This second stage is exponential, with a rate of $0.15$--$0.17$ for $U\geq 5$, and the envelope is still $0.08$ at $t=9.5$.

The second feature is a renormalization of the oscillation frequency. The first minimum of $d(t)$ lies at about $t=\pi/U$ for all $U\geq 2.5$, for example at $t=1.056$ for $U=3$, against $\pi/U=1.047$. This does not mean that the oscillation starts at the bare frequency. The Gaussian decay of the envelope moves the minimum earlier and offsets the slower advance of the phase. Once the envelope decays more slowly, the oscillations fall behind in phase, as seen in Fig.~\figref{fig: strong}{(a)}, the more so the smaller $U$. Since the extrema are shifted by the envelope, we determine the frequency by fitting a damped cosine with a linearly drifting center,
\be
d_\mathrm{fit}(t) = \bar{d} + s\,t + A(t)\cos(\omega t+\alpha), \label{eq: fit}
\ee
with the envelope
\be
\begin{aligned}
A(t) &= a_0\big[(1-w)\,e^{-\zeta t^2} + w\,\mathrm{sech}(\gamma t)\big],\\
\sigma^2/2 &= (1-w)\,\zeta + w\gamma^2/2.
\end{aligned} \label{eq: env}
\ee
This envelope has the two stages of Fig.~\figref{fig: strong}{(b)}. The second line of Eq.~\eqref{eq: env} fixes $\zeta$ such that the initial decay of $A(t)$ is the analytic short-time result $A(t)/A(0)=1-\sigma^2t^2/2+\bigO(t^4)$, with $\sigma^2=3/4$, and at late times it approaches the exponential decay $2wa_0e^{-\gamma t}$. The fit has seven free parameters: the center $\bar d + st$ with its drift $s$, the amplitude $a_0$, the frequency $\omega$, the phase $\alpha$, and the weight $w$ and rate $\gamma$ of the slow stage. We fit $d(t)$ over the whole window $0\leq t\leq 10$. Since the envelope has the analytic short-time form, no part of the window needs to be excluded. The fit reproduces $d(t)$ to an absolute rms deviation of $0.6$--$24\times10^{-4}$, at most $2.8\%$ of the initial oscillation amplitude $|E_0|/(2U)$. The largest deviations occur for $U\leq 3.5$, where a single damped cosine describes $d(t)$ less well. For $U\geq 5$, the rms deviation is below $3\times10^{-4}$, with $w\simeq 0.23$ and $\gamma\simeq 0.19$. The local decay rate of the slow stage, $\gamma\tanh\gamma t$, is then $0.14$--$0.19$ for $5\leq t\leq 9$, consistent with the rate measured from the extrema. It gives $\omega/U=0.89$ and $0.992$ for $U=3$ and $10$, so the oscillation frequency approaches $U$ only for $U$ much larger than the bandwidth. The combination $U(U-\omega)$, shown in Fig.~\figref{fig: strong}{(c)}, decreases from $1.15$ at $U=2.5$ to $0.85$ at $U=5$ and levels off at $0.81$--$0.82$ for $U=7$--$10$. This identifies a second-order renormalization $\omega \simeq U - c/U$ of the doublon-holon excitation energy. The renormalization is beyond the strong-coupling result of Ref.~\cite{Eckstein09ThermalizationIQ}, which is of first order in the hopping and gives $\omega=U$. A frequency shift $c/U$ changes $d(t)$ by terms of order $t/U^2$, which are of second order in the hopping. To analyze the renormalization, we extended that expansion to second order in the hopping, as described in Appendix~\ref{app: c}. This gives $c$ as a ratio of Fermi-sea expectation values, which in infinite dimensions evaluates to $c = 11/20 + 3E_0^2/2 = 0.820$, consistent with the td-gGA results.

The third feature is that the oscillations are centered on the prethermal value, not on the thermal one, as shown in Fig.~\figref{fig: strong}{(d)}. Reference~\cite{Eckstein09ThermalizationIQ} derived this prethermal plateau from the strong-coupling expansion, with $d_\mathrm{stat}$ halfway between the initial and the thermal value at first order. It also observed that the first few oscillations of the DMFT data are not centered on the thermal value. The td-gGA dynamics shows that this persists over the extended time window and up to $U=10$. For $U\geq 2$, the late level of $d(t)$ lies far above $d_\infty$ and approaches $d_\mathrm{stat}$ as $U$ grows. At $U=10$ it is $0.2287$, against $d_\mathrm{stat}=0.2288$ and $d_\infty = 0.2084$. Reaching the thermal value requires the decay of doublons, i.e., the conversion of interaction energy of order $U$ into kinetic energy. This needs about $U/W$ scattering processes and is exponentially slow~\cite{Strohmaier2010doublon, Sensarma2010lifetime, Eckstein2011pump}. At intermediate coupling, where relaxation is not yet governed by doublon decay, the center still drifts toward $d_\infty$. The fitted drift $s$ of Eq.~\eqref{eq: fit} is $-2.4\times10^{-3}$ and $-1.2\times10^{-3}$ per unit time for $U=2.5$ and $3$. For $U\geq 4$, $|s|$ is below the accuracy of the fit.

\section{Conclusions and outlook}\label{sec: conclusions}
We have introduced adaptive td-gGA for the real-time dynamics of correlated electron systems. It inserts ghost orbitals during the dynamics whenever the correlations that build up require them, instead of fixing their number in advance. The dynamics can therefore start from an initial state with few ghost orbitals and avoid large-$B$ equilibria. The cost at each time is set by the ghost orbitals needed up to that time, not by the largest $B$ of the run. Adaptive td-gGA is therefore several to more than ten times more efficient than td-gGA with a fixed number of ghost orbitals. The method also carries its own error estimate. Runs with a smaller upper bound $B_\mathrm{max}$ serve as a convergence check, and the time at which the last allowed ghost pair fills gives the convergence time $t_\mathrm{conv}$ from a single run. As a demonstration, we studied interaction quenches of the half-filled Hubbard model on the Bethe lattice, starting exactly from the Fermi sea. With up to $B=17$ orbitals per spin, 16 of them ghost orbitals, the dynamics is converged in $B$ to an absolute error of $10^{-3}$ in the double occupancy over the whole range $U=0.25$--$10$ up to $t\simeq 9$ or longer. Where nonequilibrium DMFT data are available, adaptive td-gGA reproduces them within their accuracy, from the weak-coupling prethermal regime through the dynamical crossover to collapse and revival. Beyond the DMFT data, adaptive td-gGA provides converged dynamics up to three and a half times as long for $U=2.5$--$4$, and for $U>4$, beyond the parameter range covered by Ref.~\cite{Eckstein09ThermalizationIQ}.

The converged dynamics gives three physical results. First, near the crossover the double occupancy relaxes to the thermal gGA value, which reproduces the DMFT thermal values. Second, resolving the dynamics in the number of ghost orbitals shows that the sharp dynamical transition of td-GA becomes a smooth crossover. Its fastest thermalization occurs at $U\simeq 1.55$--$1.6$, within the range $1.5$--$1.65$ to which DMFT restricts the crossover. Third, at strong coupling the oscillation envelope becomes independent of $U$ for $U\geq 6$, with an initial decay set by the width of the doublon-holon band. The oscillation frequency is renormalized as $\omega\simeq U - c/U$. The td-gGA coefficient $U(U-\omega)$ is consistent with the second-order strong-coupling value $c=0.820$. For $U\geq 4$ the oscillation center shows no drift toward the thermal value within the time window, and it approaches the prethermal value as $U$ grows.

The td-gGA approach is variational and wave-function based: the state stays pure, and the quasiparticle occupation spectra are constants of motion. It nevertheless reproduces the prethermalization, the collapse and revival, and the thermalization of local observables found in DMFT. The dephasing channels relevant for these observables are thus carried by a small number of ghost orbitals, added one pair at a time as the dynamics explores them. In the adaptive scheme, the time at which a new ghost pair becomes active marks the onset of a new dephasing channel. For a given $U$ these times are nearly equally spaced, $1.5$ time units apart at $U=1.5$ and $1.2$ at $U=2.5$. The thermal end point itself is compact: thermal gGA is converged already at $B=3$, so its embedding state can be purified~\cite{Verstraete2004MatrixPD} with $2(B+1)=8$ orbitals per spin, that is, the impurity and $2B+1=7$ bath orbitals, the size of a td-gGA run with $B=7$. The growth of $B$ with time thus reflects the memory that exact propagation retains, not the complexity of the state that local observables approach. In nonequilibrium DMFT, truncating the memory time of the Kadanoff-Baym equations gives access to long times~\cite{Schuler2018memory}. Whether a controlled truncation of this memory can likewise bound $B$ at long times remains to be explored. The ghost orbitals also capture the second-order renormalization of the doublon-holon excitation energy. As a result, one method, with one convergence check in $B$, describes the weak-coupling prethermal regime, the crossover and the collapse-and-revival regime, with no change of solver or of approximation across the crossover.

We see four directions for future work. First, since the required $B$ grows about linearly with time for the quench studied here, the embedding problem, a single impurity with $B$ bath orbitals per spin, is the natural target for tensor-network~\cite{Wolf2014MPSnoneqDMFT, Thoenniss2023impurity, Ng2023influence} and quantum solvers. These would make larger $B$ and longer times accessible. The Gutzwiller quantum-classical embedding (GQCE) framework~\cite{gqce, chen2025quantum} solves the embedding problem on a quantum processor and the quasiparticle problem on classical hardware. Its time-dependent extension, t-GQCE~\cite{Aftergood2026tGQCE}, is a general framework for dynamics simulations of multi-band systems based on multi-band td-gGA. There the embedding state is propagated on the quantum processor, for instance with AVQDS. Adaptive td-gGA fits this division of labor. Each ghost pair adds four qubits only once the dynamics requires it, and the population $n_\mathrm{g}$ signals in advance when this happens. In the same way, AVQDS extends its circuit only when the variational error requires it. The converged td-gGA dynamics of this quench, with its built-in error estimate, provides a reference for such solvers. Second, longer times at strong coupling would allow a quantitative test of the exponentially slow decay of doublons. Third, the scheme can be extended to multi-orbital systems. There, gGA improves substantially on the Gutzwiller approximation in equilibrium~\cite{MejutoZaera2023screening, Lee2023multiorbGRISB}, td-GA has been formulated for multi-band models~\cite{Oelsen2011multiband}, and td-gGA for multi-band systems within t-GQCE~\cite{Aftergood2026tGQCE}. However, the embedding problem of a site with several correlated orbitals is much larger. A fixed number of ghost orbitals per orbital therefore quickly becomes prohibitive. Applying the insertion criterion separately to each orbital or symmetry channel would add ghost pairs only where correlations build up, for instance in the orbitals most affected by Hund's coupling~\cite{Georges2013Hund}. Fourth, the scheme applies to driven systems. The criterion uses only the instantaneous embedding density matrix, not the form of the perturbation. It therefore carries over unchanged to time-dependent Hamiltonians such as interaction ramps~\cite{Sandri2012ramps}, periodic driving and excitation by electric-field pulses~\cite{RMPNonequilibriumDMFT}. These are central to ultrafast experiments on correlated materials~\cite{ultrafast_approach, de2021colloquium} and to the controlled dynamics of fermionic atoms in optical lattices~\cite{Bloch2008RMP, Esslinger2010}.

\section*{Data availability}
The data that support the findings of this study and the scripts that produce its figures are openly available in Zenodo at \url{https://doi.org/10.5281/zenodo.24484088}.

\section*{Acknowledgment}
The author is grateful to Nicola Lanat\`a, Daniele Guerci, Samuele Giuli, Olivier Gingras, Tsung-Han Lee, Feng Zhang, Martin Mootz, and Cai-Zhuang Wang for discussions. This work was supported by the U.S. Department of Energy (DOE), Office of Science, Basic Energy Sciences, Materials Science and Engineering Division, including the grant of computer time at the National Energy Research Scientific Computing Center (NERSC) in Berkeley, California. The research was performed at the Ames National Laboratory, which is operated for the U.S. DOE by Iowa State University under Contract No. DE-AC02-07CH11358.
\appendix
\section{Population of a newly inserted ghost pair}\label{app: ng}
Here we derive the short-time results of Sec.~\ref{sec: adaptive} for the first insertion, at $t=0$ into the Fermi sea, which $B=1$ describes exactly. The inserted pair consists of a nearly empty orbital $a$ and a nearly filled orbital $b$, related by particle-hole symmetry. The original orbital is labeled $1$.

We first consider the exact dynamics. For short times, the Heisenberg equation of motion gives $\ca_\sigma(t)=\ca_\sigma+it[\h,\ca_\sigma]+\bigO(t^2)$~\cite{Sakurai2011}. The hopping term of Eq.~\eqref{eq: hubbard} gives a one-body operator, a combination of $\ca_{J\sigma}$ on the neighboring sites. The interaction gives $[U(\hat{n}_{\up}-\tfrac12)(\hat{n}_{\dw}-\tfrac12),\ca_\sigma]=-U(\hat{n}_{\sbar}-\tfrac12)\ca_\sigma$ on each site. Hence
\be
\ca_\sigma(t) = \ca_\sigma - \frac{iUt}{2}\,\hat{X}_\sigma + \dots,\qquad \hat{X}_\sigma = 2\left(\hat{n}_{\sbar}-\tfrac12\right)\ca_\sigma ,
\ee
where the dots stand for the one-body hopping term and for higher orders in $t$. The composite operator is the difference $\hat{n}_{\sbar}\ca_\sigma-(1-\hat{n}_{\sbar})\ca_\sigma$ of the two Hubbard operators that describe the upper and lower Hubbard bands~\cite{Hubbard1963}. Since $(\hat{n}_{\sbar}-\frac12)^2=\frac14$, the composite operator is normalized, $\{\hat{X}_\sigma,\hat{X}_\sigma^\dagger\}=1$. At half filling it is orthogonal to the single-particle operator, $\braket{\{\hat{X}_\sigma,\cc_\sigma\}}_0=2\braket{\hat{n}_{\sbar}-\frac12}_0=0$. The quench thus transfers the local spectral weight $U^2t^2/4$ from $\ca_\sigma$ to $\hat{X}_\sigma$, in the spirit of the moment analysis of the Hubbard bands of Ref.~\cite{HarrisLange1967}.

In td-gGA the physical operator acts on the quasiparticles through the renormalization vector, $\cc_\sigma\to\sum_a\R_a\fc_{a\sigma}$, with the sum rule $\R^\dag\R=1$~\cite{ga_pu, gga_lanata2017, Lanata2022operatorial, Guerci2023TDgGA}. The composite operator has no counterpart in the space of the original orbital, so its weight is carried by the new pair. Particle-hole symmetry splits it equally between the two orbitals of the pair, $|\R_a|^2=|\R_b|^2=U^2t^2/8$, and leaves $|\R_1|^2=1-U^2t^2/4$. These are the values found in the simulations, as shown in Fig.~\figref{fig: ng}{(a)} below. The vector $\R$ follows by inverting the relation $\avihi{\Phi}{\cc_\sigma\ba_{a\sigma}} = [\R^\mathrm{T}\sqrt{\Delta(1-\Delta)}]_a$ of Sec.~\ref{sec: tdgga}~\cite{ga_pu, Lanata2022operatorial}. Its component along an eigenvector of $\Delta$ with a small eigenvalue is the ratio of a small embedding amplitude and the square root of that eigenvalue. It is therefore finite however small the occupation of the new pair is.

The quasiparticles evolve by Eq.~\eqref{eq: eom_n} with $M=(\R\R^\dag)^\mathrm{T}$, whose element $M_{1a}=\R_a\R_1^*$ has magnitude $Ut/(2\sqrt{2})$ to leading order. At $t=0$, $\nu(\eps)$ is diagonal, with $\nu_{11}=n_0(\eps)=\theta(-\eps)$, $\nu_{aa}=0$ and $\nu_{bb}=1$, where $\theta$ is the Heaviside step function. To first order in $M$, as in time-dependent perturbation theory~\cite{Sakurai2011},
\be
\nu_{1a}(\eps,t) = -i\eps\left[\nu_{11}-\nu_{aa}\right]\!\int_0^t\! M_{1a}(t')\,dt' = -i\eps\,n_0(\eps)\,\beta(t),
\ee
with $|\beta(t)|=Ut^2/(4\sqrt{2})$. Because Eq.~\eqref{eq: eom_n} rotates $\nu(\eps)$ unitarily, $\nu(\eps)$ remains a projector, $\nu^2=\nu$, as the one-body density matrix of a Slater determinant~\cite{Lowdin1955}. Its diagonal element $\nu_{aa}=\sum_k\nu_{ak}\nu_{ka}$ is therefore $|\nu_{1a}|^2$ up to terms of higher order, so that
\be
\nu_{aa}(\eps,t) = \eps^2\,n_0(\eps)\,|\beta(t)|^2 = \frac{U^2t^4}{32}\,\eps^2\,n_0(\eps).
\ee
Only occupied states, $\eps<0$, are transferred, with a weight proportional to $\eps^2$. The empty states fill the hole in orbital $b$ in the same way.

The embedding density matrix equals the band integral of the quasiparticle one, $\Delta=\int d\eps\,\rho_0(\eps)\,\nu(\eps)$. This is one of the gGA constraints~\cite{ga_pu, gga_lanata2017, Lanata2022operatorial}. In the block of orbitals $1$ and $a$, it gives $\Delta_{11}=\frac12$ and $\Delta_{aa}=\frac{U^2t^4}{32}\int_{-1}^{0}d\eps\,\rho_0\,\eps^2$. The off-diagonal element is $\Delta_{1a}=-i\beta\int_{-1}^{0}d\eps\,\rho_0\,\eps$. It describes a rotation that mixes the ghost orbital into the original one and carries no occupation of its own. The occupation of the ghost orbital, $n_\mathrm{g}$, is the smallest eigenvalue of this block. Second-order perturbation theory in $\Delta_{1a}$~\cite{Sakurai2011} gives it as
\be
n_\mathrm{g} = \Delta_{aa} - \frac{|\Delta_{1a}|^2}{\Delta_{11}-\Delta_{aa}} + \dots = \Delta_{aa} - 2|\Delta_{1a}|^2 + \dots,
\ee
which is Eq.~\eqref{eq: ng}. For the semicircular density of states of the Bethe lattice~\cite{dmft_georges96}, $\int_{-1}^{0}d\eps\,\rho_0\,\eps^2=\frac18$ and $\int_{-1}^{0}d\eps\,\rho_0\,\eps=-\frac{2}{3\pi}$, so that $n_\mathrm{g}=CU^2t^4$ with $C=\frac{1}{256}-\frac{1}{36\pi^2}=1.092\times10^{-3}$. The seed weight $\delta$ does not enter. It only fixes the orbitals of the new pair and makes $\R_a$ well defined.

\begin{figure}[t]
	\centering
	\includegraphics[width=\columnwidth]{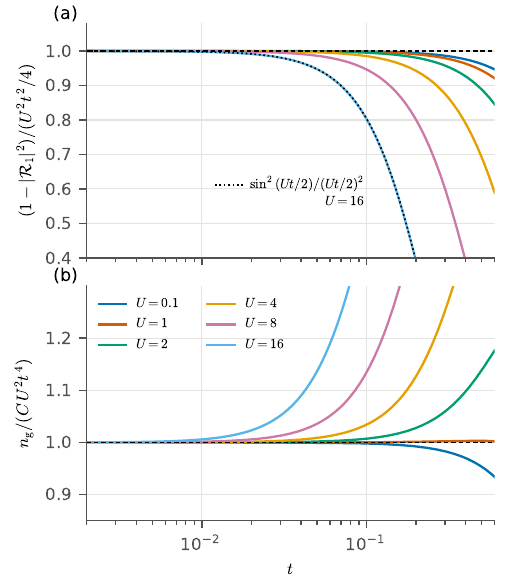}
	\caption{Short-time dynamics of the ghost pair inserted at $t=0$ into the Fermi sea for several final interactions $U$. (a) Weight moved off the original orbital divided by its leading-order value $U^2t^2/4$. The dotted line is the atomic limit $\sin^2(Ut/2)/(Ut/2)^2$ for $U=16$. (b) Population of the pair divided by the leading-order result $CU^2t^4$ of Eq.~\eqref{eq: ng}, with $C=\frac{1}{256}-\frac{1}{36\pi^2}$. The dashed lines mark the ratio $1$. The td-gGA runs use $B=3$. They are averaged over two opposite signs of the seed excitation, which removes the contribution of the seed to the population. This contribution is linear in the seed amplitude, $10^{-6}$, and is visible in the ratio only at small $Ut^2$.}
	\label{fig: ng}
\end{figure}

Figure~\ref{fig: ng} compares these results with td-gGA. We insert the first ghost pair at $t=0$ and propagate with $B=3$ for twelve final interactions from $U=0.1$ to $16$. The seed weight is $\delta=10^{-12}$, smaller than the $\delta=10^{-8}$ used for the results of Sec.~\ref{sec: results}, so that the seed contribution stays small even at the shortest times. The figure shows six of them. The seed adds to $n_\mathrm{g}$ a cross term between its amplitude $\sqrt{\delta}$ and the driven amplitude $\sqrt{C}\,Ut^2$, so its relative size is $2\sqrt{\delta}/(\sqrt{C}\,Ut^2)\approx60\sqrt{\delta}/(Ut^2)$. We remove it by averaging two runs with opposite signs of the seed excitation. Figure~\figref{fig: ng}{(a)} first confirms the weights entering the derivation. The weight moved off the original orbital, $1-|\R_1|^2$, does not depend on the seed and approaches $U^2t^2/4$ to a relative error of $10^{-4}$ at all $U$. It is shared equally by the two new orbitals, $|\R_a|^2=|\R_b|^2$ to all printed digits, and the sum rule holds to $5\times10^{-5}$ up to $t=0.1$. Its departure from the leading order follows the atomic limit. The interaction alone gives $\ca_\sigma(t)=\ca_\sigma\cos(Ut/2)-i\hat{X}_\sigma\sin(Ut/2)$ exactly, which moves the weight $\sin^2(Ut/2)$. The td-gGA weight follows it to a relative error of $1.5\times10^{-3}$ up to $t=0.1$ at all $U$. In Fig.~\figref{fig: ng}{(b)}, the ratio $n_\mathrm{g}/(CU^2t^4)$ approaches $1$ for $t\to0$ at all $U$. A fit of its short-time limit gives $1$ to within $5\times10^{-5}$ at all $U$. Equivalently, the computed coefficient agrees with the analytical $C$ to a relative accuracy of $5\times10^{-5}$. Since the two terms of $C$ nearly cancel, this agreement tests the rotation term as well as the transferred occupation. The curves depart from $1$ at a time set by the next order of the expansion. Equation~\eqref{eq: ng} holds to $2\%$ for $t\lesssim0.1$ and $Ut\lesssim0.3$. At the first insertion, where $n_\mathrm{g}$ reaches $n_\mathrm{ins}=10^{-5}$, the ratio lies between $0.97$ and $1.06$ for $U=0.5$ to $3$ and reaches $1.15$ at $U=8$. Since $n_\mathrm{g}$ grows as $t^4$ at these short times, a ratio of $1.15$ shifts the insertion time earlier by $1-1.15^{-1/4}\approx3.4\%$, so Eq.~\eqref{eq: ng} gives the insertion time to within $4\%$ up to $U=8$.

\section{Second-order frequency of the doublon oscillation}\label{app: c}
Reference~\cite{Eckstein09ThermalizationIQ} derives the strong-coupling dynamics after the quench to first order in the hopping. At this order the double occupancy oscillates with angular frequency $U$. Here we extend that derivation to second order, which gives the renormalized angular frequency $\Omega=U-c/U$ with which the fitted $\omega$ of Sec.~\ref{sec: strong} is compared. We summarize the steps necessary for the derivation. Up to conserved terms, which do not affect the dynamics, the Hamiltonian~\eqref{eq: hubbard} is $\hat{H}=U\hat{D}+\hat{T}$, where $\hat{D}=\sum_I \hat{n}_{I\up}\hat{n}_{I\dw}$ counts the doubly occupied sites and $\hat{T}$ is the hopping term. Following Refs.~\cite{HarrisLange1967, Eckstein09ThermalizationIQ}, we sort each hopping process $t_\mathrm{h}\,\cc_{I\sigma}\ca_{J\sigma}$ by the occupation of the opposite spin $\sbar$ on the two sites, $\hat{T}=\hat{T}_0+\hat{T}_++\hat{T}_-$, with
\begin{align}
\hat{T}_+ &= t_\mathrm{h}\sum_{IJ\sigma} \hat{n}_{I\sbar}\,\cc_{I\sigma}\ca_{J\sigma}\,(1-\hat{n}_{J\sbar}), \nonumber\\
\hat{T}_- &= t_\mathrm{h}\sum_{IJ\sigma} (1-\hat{n}_{I\sbar})\,\cc_{I\sigma}\ca_{J\sigma}\,\hat{n}_{J\sbar}, \\
\hat{T}_0 &= t_\mathrm{h}\sum_{IJ\sigma} \left[\hat{n}_{I\sbar}\hat{n}_{J\sbar} + (1-\hat{n}_{I\sbar})(1-\hat{n}_{J\sbar})\right]\cc_{I\sigma}\ca_{J\sigma}, \nonumber
\end{align}
where the sums run over both orders of each nearest-neighbor pair. Here $\hat{T}_+$ creates a doublon-holon pair, $\hat{T}_-=\hat{T}_+^\dagger$ annihilates one, and $\hat{T}_0$ conserves $\hat{D}$, which is expressed by the identity
\begin{equation}
[\hat{D},\hat{T}_m]=m\hat{T}_m,\qquad m=0,\pm1.
\label{eq: DT}
\end{equation}

As in the derivation of the $t$-$J$ model~\cite{Chao1977kinetic, MacDonald1988tU}, we use a Schrieffer-Wolff transformation~\cite{SchriefferWolff1966, Bravyi2011SW}. This unitary transformation $e^{\hat{S}}$, with $\hat{S}^\dagger=-\hat{S}$, removes the terms that change $\hat{D}$ from $\hat{H}_\mathrm{eff}=e^{\hat{S}}\hat{H}e^{-\hat{S}}$ order by order in $t_\mathrm{h}/U$. By Eq.~\eqref{eq: DT}, a term $\hat{O}_m$ that changes $\hat{D}$ by $m\neq0$ is removed by adding $\hat{O}_m/(mU)$ to $\hat{S}$. We apply this at first order to $\hat{T}_\pm$ and at second order to the terms $\pm[\hat{T}_\pm,\hat{T}_0]/U$ that the first step generates. This gives
\begin{equation}
\hat{S} = \frac{\hat{T}_+ - \hat{T}_-}{U} + \frac{[\hat{T}_+,\hat{T}_0] + [\hat{T}_-,\hat{T}_0]}{U^2} + \bigO(U^{-3}),
\end{equation}
and $\hat{H}_\mathrm{eff} = U\hat{D} + \hat{K}$ with $\hat{K}=\hat{T}_0 + [\hat{T}_+,\hat{T}_-]/U + \bigO(U^{-2})$, which commutes with $\hat{D}$.

The double occupancy follows from $e^{-i\hat{H}t}=e^{-\hat{S}}e^{-i\hat{H}_\mathrm{eff}t}e^{\hat{S}}$ as $d(t)=\braket{\phi|e^{i\hat{H}_\mathrm{eff}t}\hat{D}'e^{-i\hat{H}_\mathrm{eff}t}|\phi}/L$ for $L$ sites. Here $\ket{\psi_0}$ is the Fermi sea, $\ket{\phi}=e^{\hat{S}}\ket{\psi_0}$ the rotated initial state and $\hat{D}'=e^{\hat{S}}\hat{D}e^{-\hat{S}}$ the rotated observable. We sort $\hat{D}'$ by its change of $\hat{D}$. To second order, its part that conserves $\hat{D}$ is $\hat{D}'_0=\hat{D}-[\hat{T}_+,\hat{T}_-]/U^2$. Its part that raises $\hat{D}$ by one is
\begin{equation}
\hat{A} = -\frac{\hat{T}_+}{U} - \frac{[\hat{T}_+,\hat{T}_0]}{U^2} + \bigO(U^{-3}),
\end{equation}
and its part that lowers $\hat{D}$ is $\hat{A}^\dagger$. Since $\hat{K}$ commutes with $\hat{D}$, Eq.~\eqref{eq: DT} gives $e^{i\hat{H}_\mathrm{eff}t}\hat{A}e^{-i\hat{H}_\mathrm{eff}t}=e^{iUt}e^{i\hat{K}t}\hat{A}e^{-i\hat{K}t}$, so that
\begin{equation}
\begin{aligned}
d(t) = &\frac{1}{L}\braket{\phi|e^{i\hat{K}t}\hat{D}'_0e^{-i\hat{K}t}|\phi} \\ &+ \frac{2}{L}\,\mathrm{Re}\left[e^{iUt}\braket{\phi|e^{i\hat{K}t}\hat{A}e^{-i\hat{K}t}|\phi}\right].
\end{aligned}
\label{eq: dt_sc}
\end{equation}
The first term evolves only with $\hat{K}$, on the time scale of the inverse bandwidth, and is the slowly varying center of $d(t)$. The second term is the oscillation. At first order we set $\ket{\phi}\simeq\ket{\psi_0}$ and $\hat{A}\simeq -\hat{T}_+/U$, neglect the slow evolution with $\hat{K}$, and use $\braket{\hat{T}_\pm}_0/L=E_0/4$ in the Fermi sea. Equation~\eqref{eq: dt_sc} then reduces to Eq.~\eqref{eq: d1}, the result of Ref.~\cite{Eckstein09ThermalizationIQ}.

Beyond first order, the evolution with $\hat{K}$ spreads the oscillation over a band of frequencies. In the eigenbasis of $\hat{H}_\mathrm{eff}$, $\braket{\phi|e^{i\hat{H}_\mathrm{eff}t}\hat{A}e^{-i\hat{H}_\mathrm{eff}t}|\phi}=\sum_{mn}\phi_m^*A_{mn}\phi_n e^{i(E_m-E_n)t}$. Each frequency $E_m-E_n$ is $U$ plus a difference of eigenvalues of $\hat{K}$. This band of the width of the kinetic energy dephases and damps the oscillation. Its centroid, the weighted mean of $E_m-E_n$, is the first moment, as in the moment analysis of the Hubbard bands of Ref.~\cite{HarrisLange1967}:
\begin{equation}
\Omega = \frac{\braket{\phi|[\hat{H}_\mathrm{eff},\hat{A}]|\phi}}{\braket{\phi|\hat{A}|\phi}} = U + \frac{\braket{\phi|[\hat{K},\hat{A}]|\phi}}{\braket{\phi|\hat{A}|\phi}},
\end{equation}
which requires only expectation values in $\ket{\phi}$. The centroid is the angular frequency with which the oscillation starts, $\Omega=U-i\dot{F}(0)/F(0)$ with $F(t)=\braket{\phi|e^{i\hat{K}t}\hat{A}e^{-i\hat{K}t}|\phi}$. At later times the phase velocity of $F(t)$ depends also on the shape of the band.

We expand $\Omega$ with $\braket{\phi|\hat{O}|\phi}=\braket{\hat{O}}_0+\braket{[\hat{O},\hat{S}]}_0+\dots$, where $\braket{\cdot}_0$ is the expectation value in the Fermi sea. The correction of order $t_\mathrm{h}$, $\braket{[\hat{T}_0,\hat{T}_+]}_0/\braket{\hat{T}_+}_0$, vanishes by the particle-hole symmetry of one spin species. This symmetry leaves $\hat{T}_0$ and the half-filled Fermi sea invariant and exchanges $\hat{T}_+$ and $\hat{T}_-$. The terms of order $t_\mathrm{h}^2/U^2$ have three sources: the virtual creation and annihilation of a doublon-holon pair in $\hat{K}$, the motion of a pair by one hop right after its creation in $\hat{A}$, and the rotation of the initial state. Together they give
\begin{equation}
\Omega = U - \frac{c}{U},\qquad c = -\frac{\braket{\hat{Q}}_0}{\braket{\hat{T}_+}_0},
\label{eq: c}
\end{equation}
with
\begin{equation}
\hat{Q} = [\hat{T}_0,[\hat{T}_+,\hat{T}_0]] + [[\hat{T}_0,\hat{T}_+],\hat{T}_+ - \hat{T}_-] + [[\hat{T}_+,\hat{T}_-],\hat{T}_+].
\end{equation}
Exact diagonalization of the quench on rings of eight sites confirms Eq.~\eqref{eq: c}: $U(U-\Omega)$ of the exact spectral centroid approaches $c$ to $1\times10^{-3}$ at $U=160$ in units of the hopping.

Each term of $\braket{\hat{Q}}_0$ is a product of three hopping processes on at most four sites. Since the Fermi sea is a Gaussian state~\cite{Peschel2003RDM}, each term follows from Wick's theorem in the single-particle correlations $\braket{\cc_{I\sigma}\ca_{J\sigma}}_0$. We take the Bethe lattice with coordination $z$ and $t_\mathrm{h}=1/(2\sqrt{z})$, which keeps the band edges at $\pm1$. The correlations at distance one and three are then $-g_1/\sqrt{z}$ and $-g_3/z^{3/2}$, with $g_1=|E_0|=4/(3\pi)$ and $g_3=-8/(15\pi)$ for the semicircular density of states. Because $\braket{\hat{Q}}_0$ is odd in the correlations at odd distance, only the terms linear in $g_1$ and $g_3$ and cubic in $g_1$ survive for $z\to\infty$. Counting the hopping triples exactly as polynomials in $z$ and using $\braket{\hat{T}_+}_0/L=-g_1/4$ gives
\begin{equation}
c = \frac{1}{2} + \frac{3}{2}g_1^2 - \frac{g_3}{8g_1} = \frac{11}{20} + \frac{3}{2}E_0^2 = 0.8202,
\end{equation}
in agreement with a numerical evaluation at $z=3$ to $32$ extrapolated in $1/z$. A fit of $d(t)$ with a single frequency averages the phase velocity over the fit window, weighted by the oscillation amplitude. This amplitude decays within a few inverse bandwidths, so that the fit is dominated by early times.

The same expansion gives the decay of the oscillation at short times. At first order, with $\ket{\phi}\simeq\ket{\psi_0}$, $\hat{A}\simeq-\hat{T}_+/U$ and $\hat{K}\simeq \hat{T}_0$, $F(t)$ is proportional to $F_0(t)=\braket{e^{i\hat{T}_0t}\hat{T}_+e^{-i\hat{T}_0t}}_0=\sum_n (it)^n\braket{\mathcal{L}^n \hat{T}_+}_0/n!$, where $\mathcal{L}\hat{O}=[\hat{T}_0,\hat{O}]$. The terms with odd $n$ vanish by the particle-hole symmetry that removes $\braket{[\hat{T}_0,\hat{T}_+]}_0$, so that
\begin{equation}
\frac{F_0(t)}{F_0(0)} = 1-\frac{\sigma^2t^2}{2}+\bigO(t^4),\qquad \sigma^2=\frac{\braket{[\hat{T}_0,[\hat{T}_0,\hat{T}_+]]}_0}{\braket{\hat{T}_+}_0},
\end{equation}
where $\sigma^2$ is the variance of the band. Evaluating the double commutator in the same way as $\braket{\hat{Q}}_0$ gives $\braket{[\hat{T}_0,[\hat{T}_0,\hat{T}_+]]}_0/L=-3g_1/16$ for $z\to\infty$. With $\braket{\hat{T}_+}_0/L=-g_1/4$, we thus have $\sigma^2=3/4$.

\bibliography{main}

\end{document}